\documentclass[aps,pra,twocolumn,showpacs,groupedaddress,nofootinbib]{revtex4}
\usepackage{graphicx}
\usepackage{hyperref}
\usepackage{color}

\begin{document}

\title{Least Action Principle in Gait}

\author{Yifang Fan$^{a}$\footnote{Corresponding author\\ Email: tfyf@mailer.gipe.edu.cn}, Mushtaq Loan$^{b}$,
 Yubo Fan$^{c}$, Zongxiang Xu$^{a}$, Zhiyu Li$^{d}$ and Donglin Luo$^{a}$}
\affiliation{$^{a}$Center for Scientific Research, Guangzhou
Institute of Physical Education, Guangzhou 510500, China\\
$^{b}$International School, Jinan University,Guangzhou 510632, China\\
$^{c}$Bioengineering Department, Beijing University of
Aeronautics and Astronautics, Beijing 100191, China\\
$^{d}$College of Foreign Languages, Jinan University, Guangzhou
510632, China}

\date{\today}

\begin{abstract}
We apply the laws of human gait vertical ground reaction force and
discover the existence of the phenomenon of least action principle
in gait. Using a capacitive mat transducer system, we obtain the
variations of human gait vertical ground reaction force and
establish a structure equation for the resultant of such a force.
Defining the deviation of vertical force as an action function, we
observe from our  gait optimization analysis the least action
principle at half of the stride time. We develop an evaluation index
of mechanical energy consumption based upon the least action
principle in gait. We conclude that these observations can be
employed to enhance the accountability of gait evaluation.
\end{abstract}


\maketitle

%\section{INTRODUCTION}

Gait analysis explores laws of body movement by bio-mechanical
methods so that it can serve the clinical diagnosis and
rehabilitation. Gait parameters refer to physical quantities while
walking, for example, the space-time characteristics, kinesiological
quantities and kinetic quantities \cite{vaughan,saunders}. The
relative symmetry of gait parameters is a notable feature of normal
human natural gait \cite{mitoma,kim}. Phase symmetry index was
proposed as gait quantization index and mathematical model and it
has brought satisfactory result in the study of walking function of
hemiplegic patients \cite{dewar,waii,ekaterina}. With the
development of gait measurement equipment, more gait evaluation
index of gait parameters have been defined
\cite{kondraske,wang,ajoy}. The essence of gait index evaluation is
to compare, that is, to take the normal human gait as the standard
of rehabilitation evaluation. However, the non-complete symmetry of
human shape, coordinate and strength has formed the uniqueness of
human gait \cite{murray}, which has brought difficulty to the
establishment of gait parameter index standard. Consequently,
exploring a law that is irrelevant to one's age, gender or physical
shape, in gait has become very important when applying gait to
scientifically evaluate the rehabilitation.

Bipedal walking has enabled the continuous evolution of human gait
\cite{jenkins}, which eventually brought about the optimized gait
\cite{srinivasan} and formed a human behavioral trait. Nature always
minimizes certain important quantities when a physical process takes
place \cite{marion}. In the continuous evolution of this human
behavioral trait, explanations to issues such as how it observes the
least action principle (hereinafter referred to as LAP) and what the
action function in gait is remain controversial. In this letter we
address the issue of the principles of least action in gait and
propose a more reliable and standard gait evaluation index system.

Gait is under the control of the nervous system. Its musculoskeletal
system generates resultant force, which acts on the ground by foot,
and in turn, the consequent vertical ground reaction force (VGRF)
enables the human body to move \cite{winter}. The vertical change of
mass center can be analyzed either by the space-time relation of
human segments or by VGRF \cite{crowe}.

\begin{figure}[!h]
\scalebox{0.7}{\includegraphics{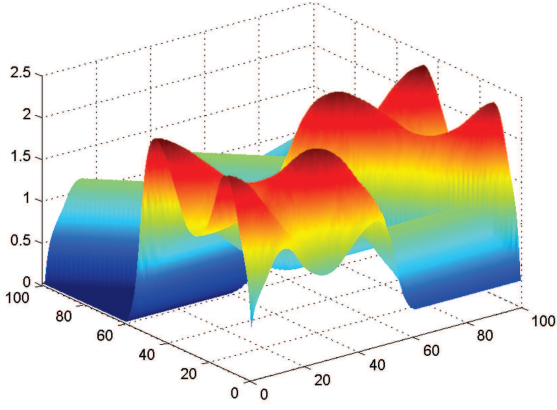}} \caption{ \label{fig1}
Relationship between the landing sequences of the foot and its
variations of VGRF. The variation range of both $t$ and $t_{0}$ is
defined over the range $[0,T]$.}
\end{figure}

Following the human gait characteristics, the equation of certain
moment's VGRF resultant force\footnote{We have assumed that the VGRF
of both left and right feet are the same.} is
\begin{equation}
F_{GRF}(t)=F_{z}^{r}(t)+F_{z}^{l}(t), \label{eqn1}
\end{equation}
where $F_{z}^{r}(t)=F_{z}(t)$  and $F_{z}^{l}(t)=F_{z}(t+t_{0})$,
for $0\leq t\leq T$, are the VGRFs of right and left feet,
respectively and $T$ is one foot stride time. Using Eq. \ref{eqn1},
one can obtain the landing sequences of the foot and its variations
of VGRF. Rating the cycle time as a percentage and normalizing the
reaction force by $F_{z}(t)/mg$ and $F_{z}(t+t_{0})/mg$, we display
the relationship between the variation of VGRF and the landing
sequences of the foot in Fig.\ref{fig1}.

Analyzing the spectrum in Fig.\ref{fig1}, we notice, after
determining each foot's variation of vertical ground reaction force,
that $\frac{1}{Tmg}\int_{0}^{T}F_{GRF}(t)dt-1=0$ in each stride
cycle. In this way, the change of  $t_{0}$ leads to the change of
the distribution of VGRF in a stride cycle. The distribution of
$F_{GRF}$  actually determines the walking movement. For example,
when $t_{0}=0$, the walking will become jumping. In order to analyze
the influence of foot land sequence to the distribution of VGRF
resultant, we turn to the concept of deviation distribution.
Using the deviation of VGRF resultant force, $\sqrt{\frac{\sum
\big(F_{z}(t)/mg+F_{z}(t+t_{0})/mg\big)^{2}-Tf}{Tf-1}}$($f$ being
the collection frequency of the measurement system), we develop the
relationship between the landing sequence of one foot and its
deviation of VGRF resultant force. Using the optimization method, we
found that the action function $\Psi (t_{0})$ $\big(=
\sum_{t=1/f}^{T} \big(F_{Z}(t)+F_{Z}(t+t_{0})-mg\big)^{2}\big)$, of
the resultant of VGRF in a gait cycle reaches an optimum value
around $\frac{1}{2}T$.
\begin{figure}[!h]
\scalebox{0.70}{\includegraphics{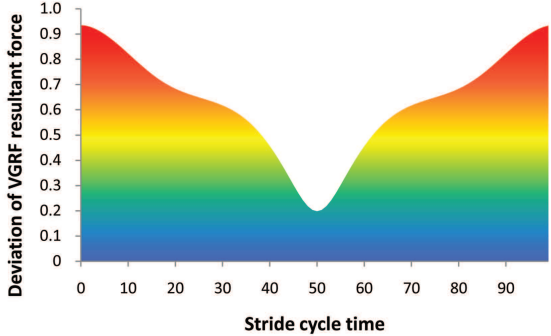}} \caption{ \label{fig2} Plot
showing the deviation of resultant vertical ground reaction force as
a function of stride cycle time. The stride time is rated as
percentage by $(t/T)\times 100$.}
\end{figure}
As can be seen in Fig. \ref{fig2}, this change is symmetric and has
a minimal value about $\frac{1}{2}T$. Therefore we conclude that in
gait, when the starting time of one foot stride cycle time falls
right at the half of the other foot stride time, the deviation of
VGRF is the minimal.

The movement of human segment is done by a combination of prime
movers, antagonist, and synergists. The extension and bending of a
segment's movement consume mechanical energy. Since in a stride
cycle, $E_{g}=0$ and $E_{k}= \frac{1}{Tf}\sum \frac{1}{2}mv_{t}^{2}
>0 $ thus we use $E_{k}$ to describe the vertical mechanical energy (VME)
consumption in gait.
\begin{figure}[!h]
\scalebox{0.70}{\includegraphics{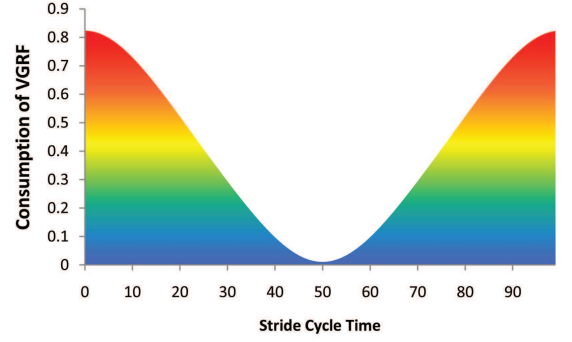}} \caption{ \label{fig3}
Consumption of vertical mechanical energy as a function of stride
cycle time.}
\end{figure}
Fig. \ref{fig3} illustrates effect the landing sequences of the foot
impose upon the variation of vertical mechanical energy. It follows
that while walking, one rearfoot touches the ground exactly at the
half of the other foot's stride time and consecutive gaits consume
the least VME. This is so-called the least action principle in gait
(LAPG).

Using the variation of VGRF to evaluate the consumption of VME in
gait, we propose an energy consumption index of the following form:
\begin{equation}
ID_{E}=\frac{\sum^{n}_{i=1}T_{i}\int^{T}_{0}|F_{z}(t)+F_{z}(t+t_{o})-mg|dt}{T\int^{T_{1}+\cdots+T_{n}}_{0}|F_{GRF}(t)-mg|dt},
\label{eqn2}
\end{equation}
where $n$ is the number of stride cycles. For $n=3$, the above
definition reproduces Zebris FDM Gait Analysis System. The advantage
with the above definition is that one does not need to calculate
vertical accelerated velocity, velocity and displacement of mass
center.

\begin{figure}[!h]
\scalebox{0.70}{\includegraphics{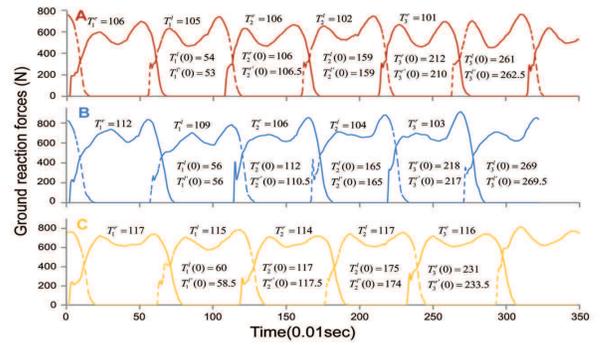}} \caption{ \label{fig4}
Shown are the plots of (A) feet pressure of a slightly injured
adult, (B) feet pressure of an adult and (C) feet pressure of an
elderly with rheumatoid arthritis.}
\end{figure}

Using the Zebris FDM-System Measurement System for Gait Analysis, we
collect the data that best represent the usual normal gaits and meet
the essential requirements of the test. We test the gait of one male
adult with slight injury in his left ankle, a second male and one
elderly male with rheumatic arthritis. The analysed results are
displayed in Fig. \ref{fig4}.
We notice that when a fit adult walks, the starting time of one foot
in a stride cycle will spontaneously fall right at the half of the
other foot's stride cycle time. Applying the VME to evaluate the
energy consumption in gait for these subjects, we obtain the
estimates of $0.931, 0.998$ and $0.743$, respectively. A comparison
with the earlier results \cite{wang} shows that the established
evaluation index can provide better accountability to different
gaits.

In order to examine its universality, we enlarged our sample size to
$173$ subjects ($95$ male and $78$ female students). The collected
data are tabulated in Table \ref{tab1}.
\begin{table}[!h]
\caption{ \label{tab1} \mbox{Mean $\pm$ S.D} age, height and body
mass of the subjects.}
\begin{ruledtabular}
\begin{tabular}{ccccccc}
Gender & Sample size & Age  & Height  & Body mass  \\
       &             & (years)     &  (m)       &  (Kgs) \\
\hline
Male &  95 & $21.1\pm 1.31$ & $1.72\pm 0.64$ & $61.8\pm 8.3$\\
Female & 78 & $21.8\pm 1.3$ & $1.61\pm 0.58$ & $51.3\pm 7.6$
\end{tabular}
\end{ruledtabular}
\end{table}
The statistical analysis to the results for gait, using WinFDM,
yield the stride times of  $ 1.01(6)$ sec and $1.00(6)$ sec for the
left and right foot, respectively. The ratio between the starting
time calculated by LAPG and that of the tested result for the left
and right foot is estimates as $0.998(21)$  and $0.993(21)$,
respectively. The excellent agreement between the estimates confirms
the signature of the universality of LAPG.

\begin{figure}[!h]
\scalebox{0.70}{\includegraphics{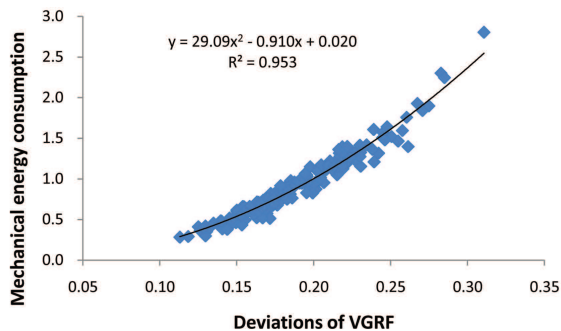}} \caption{ \label{fig5}
Consumption of VME as a function of subject's deviation of VGRF. The
solid curve is a quadratic fit to the data.}
\end{figure}

To examine the effect that the landing sequences of the foot exert
upon both the deviation of VGRF resultant and upon the consumption
of VME, we display our  proposed  relationship between the subject's
deviation of VGRF and the consumption of VME in Fig. \ref{fig5}.
An analysis of correlation reveals that the two quantities are
highly correlated ($R=0.976, p<0.001$), a least deviation of VGRF
results in least consumption of VME.

In this study, we have developed a relationship between the
deviation of VGRF and its consumption of VME and established a
description of VGRF resultant force. From our analysis we conclude
that when the deviation of VGRF in gait is least, its consumption of
VME is also the least. We have also developed an evaluation index of
mechanical energy consumption based up the LAPG. Our results
indicate the universality of least action principle in gaits.

A normal adult's gait has nothing to do with their physiology (e.g.
gender, age), body shape (e.g. height, weight) or their gait (e.g.
cadence, velocity). One foot's stride starting time always begins at
the half of the next foot's stride cycle time, which consumes the
least VME.  A regression equation of foot VGRF in a stride cycle has
been set up. Representing  the  deviation of VGRF by the action
function, we have discovered the LAPG by the optimization analysis
method. This signature was confirmed  by analyzing the consecutive
gaits of $173$ subjects. Our results suggest that the evolution of
gait, in addition to its adaptation to the natural environment
\cite{jenkins,richmond},  is a consequence of following LAP. Human
present physical condition uses the most energy-saving gait, even
after a slight injury to the ankle. This could be considered as a
human instinct. In sport rehabilitation, the therapy should be
focused on the recovery of physical function. The uniqueness of each
individual's gait is shaped when deviations of human body inertial
parameters, muscle strength, and motion coordination do exist, but
they all follow LAPG. A research into the variations of natural gait
(bare-footed) shear stress would enrich the study of LAPG. This
study has convinced us that LAP has profoundly influenced the
natural evolution, even the evolution of life. Lamarck's mechanism
for the evolution of life use and disuse is an expression of LAP. A
further research into LAPG will be significant to the studies such
as sport rehabilitation, biometric identification techniques and the
control of biped robots gaits \cite{collins,ohgane}.

This project was funded by  National Natural Science Foundation of
China under the grant $10772053$ and by Key Project of Natural
Science Research of Guangdong Higher Education Grant No $06Z019$ .
The authors would like to acknowledge the support from the subjects.

\end{document}